\documentclass[a4paper,11pt]{article}

\def\marginset#1#2{                      

\setlength{\oddsidemargin}{#1}         

\setlength{\evensidemargin}{0mm}       

\setlength{\hoffset}{\paperwidth}

\addtolength{\hoffset}{-\oddsidemargin}

\addtolength{\hoffset}{-\textwidth}

\addtolength{\hoffset}{-\evensidemargin}

\setlength{\hoffset}{0.5\hoffset}

\addtolength{\hoffset}{-1in}           

\setlength{\voffset}{-1in}             

\setlength{\topmargin}{\paperheight}

\addtolength{\topmargin}{-\headheight}

\addtolength{\topmargin}{-\headsep}

\addtolength{\topmargin}{-\textheight}

\addtolength{\topmargin}{-\footskip}

\addtolength{\topmargin}{#2}           

\setlength{\topmargin}{0.5\topmargin}

}

\marginset{0mm}{0mm}

\usepackage{times}                       

\usepackage{fancyhdr}                    

\usepackage{color}                       

\usepackage{indentfirst}                 



\usepackage{url}
\usepackage[dvips]{graphicx}
\usepackage[fleqn]{amsmath}

\begin{document}

\title{Stable Feature Selection for Biomarker Discovery}

\author{Zengyou He, Weichuan Yu \\Laboratory for Bioinformatics and Computational Biology, \\Department of Electronic and Computer
Engineering, \\The Hong Kong University of Science and Technology,
Kowloon, Hong Kong, China.}

\maketitle

\begin{abstract}

Feature selection techniques have been used as the workhorse in
biomarker discovery applications for a long time. Surprisingly, the
stability of feature selection with respect to sampling variations
has long been under-considered. It is only until recently that this
issue has received more and more attention. In this article, we
review existing stable feature selection methods for biomarker
discovery using a generic hierarchal framework. We have two
objectives: (1) providing an overview on this new yet fast growing
topic for a convenient reference; (2) categorizing existing methods
under an expandable framework for future research and development.

\emph{\textbf{Keywords:}} Feature selection; biomarker discovery;
stability; machine learning

\end{abstract}

\section{Introduction}

Recent advances in genomics and proteomics enable the discovery of
biomarkers for diagnosis and treatment of complex diseases at the
molecular level \cite{SVZ02}. A biomarker may be defined as ``a
characteristic that is objectively measured and evaluated as an
indicator of normal biological processes, pathogenic processes, or
pharmacologic responses to a therapeutic intervention''
\cite{BiomarkerDef}.

The discovery of biomarkers from high-throughput ``omics'' data is
typically modeled as selecting the most discriminating features (or
variables) for classification (e.g. discriminating healthy versus
diseased, or different tumor stages) \cite{HK08,ADW09}. In the
language of statistics and machine learning, this is often referred
to as feature selection. Feature selection has attracted strong
research interest in the past several decades. For recent reviews of
feature selection techniques used in bioinformatics, the reader is
referred to \cite{SIL07,MH08,HK08,DH09}.

While many feature selection algorithms have been proposed, they do
not necessarily identify the same candidate feature subsets if we
repeat the biomarker discovery procedure \cite{YDL08}. Even for the
same data, one may find many different subsets of features (either
from the same feature selection method or from different feature
selection methods) that can achieve the same or similar predictive
accuracy \cite{DKG05,MKH05,ZRS08}. In practice, high reproducibility
of feature selection is equally important as high classification
accuracy \cite{JMB08}. It is widely believed that a study that
cannot be repeated has little value \cite{ZZZ09}. Consequently, the
instability of feature selection results will reduce our confidence
in discovered markers.

The stability issue in feature selection has received much attention
recently. In this article, we shall review existing methods for
stable feature selection in biomarker discovery applications,
summarize them with an unified framework and provide a convenient
reference for future research and development.

This article differs from existing review papers on feature
selection in the following aspects:
\begin{itemize}
  \item Compared to current feature selection reviews \cite{SIL07,MH08,HK08,DH09}, this review focuses only on those
feature selection approaches that incorporate ``stability'' into the
algorithmic design.
  \item This article mainly focuses on ``methods''
for finding reliable markers rather than ``metrics'' of measuring
the stability of selected feature subsets \cite{BS09}, although we
also list these metrics for completeness.
\end{itemize}

The remainder of the paper is organized as follows. In section 2, we
discuss several sources that cause the instability of feature
selection. In section 3, we summarize available stable feature
selection algorithms and describe different classes of methods in
detail. In section 4, we provide a list of stability measures and
illustrate their definitions. We give a discussion in section 5.
Finally, we conclude this paper in section 6.

\section{Causes of Instability}

There are mainly three sources of instability in biomarker
discovery:
\begin{enumerate}
  \item Algorithm design without considering
  stability: Classic feature selection methods aim at selecting a minimum subset
of features to construct a classifier of the best predictive
accuracy \cite{YDL08}. They often ignore ``stability'' in the
algorithm design.
  \item The existence of multiple sets of true markers: It is possible that there exist multiple sets of potential true
markers in real data. On the one hand, when there are many highly
correlated features, different ones may be selected under different
settings \cite{YDL08}. On the other hand, even there are no
redundant features, the existence of multiple non-correlated sets of
real markers is also possible \cite{ZYG08}.
  \item Small number of samples in high dimensional data: In the analysis of gene expression data and proteomics data, there
are typically only hundreds of samples but thousands of features. It
has been experimentally verified that the relatively small number of
samples in high dimensional data is one of the main sources of the
instability problem in feature selection \cite{LYD09,Kim09}. To
understand the nature of the instability of selected feature subset,
Ein-Dor et al \cite{DZD06} developed a new mathematical model and
concluded that at least thousands of samples are needed to achieve
stable feature selection.
\end{enumerate}

Here we list three sources that can cause the instability of feature
selection in biomarker discovery. We believe that there may be still
other sources that can affect the stability of feature selection.
The identification of these sources is of primary importance for
future research and development. On the one hand, knowing the reason
enables us to better understand the problem. On the other hand, such
knowledge will facilitate the design of new methods for stable
biomarker discovery.

\section{Existing Methods}

\begin{figure}
\centerline{\includegraphics[width=0.75\textwidth]{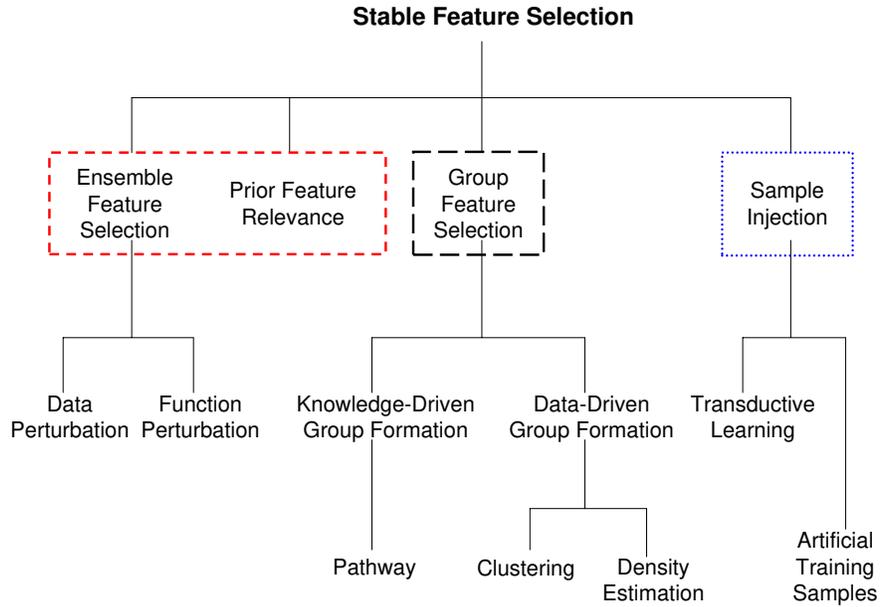}}
\caption{A hierarchical framework for stable feature selection
methods.} \label{fig:framework}
\end{figure}

To date, there are many methods available for stable feature
selection. We wish to cover all existing methods in a systematic and
expandable manner. Fig.\ref{fig:framework} illustrates our approach
to summarizing different methods based on the way they treat
different sources of instabilities. Briefly, the ensemble feature
selection method and the method using prior feature relevance
incorporate stability consideration into the algorithm design stage.
To handle data with highly correlated features, the group feature
selection approach treats feature cluster as the basic unit in the
selection process to increase robustness. The sample injection
method tries to increase the sample size to address the
small-sample-size vs. large-feature-size issue. In the following
sections, we will discuss each category in detail.

\subsection{Ensemble Feature Selection}

In statistics and machine learning, ensemble learning methods
combine multiple learned models under the assumption that ``two (or
more) heads are better than one''. Typical ensemble learning methods
such as bagging \cite{Bagging} and boosting \cite{Boosting} have
been widely used in classification and regression. Ensemble feature
selection techniques use a two-step procedure that is similar to
ensemble classification and regression:
\begin{enumerate}
  \item Creating a set of different feature selectors.
  \item Aggregating the results of component selectors to generate the ensemble
  output.
\end{enumerate}

The second step is typically modeled as a rank aggregation problem.
Rank aggregation combines multiple rankings into a consensus
ranking, which has been widely studied in the context of web search
engines \cite{DKN01}. In most cases, the strategies rely on the
following information:
\begin{itemize}
  \item The ordinal rank associated with each feature.
  \item The score assigned to each feature.
\end{itemize}

To date, many rank aggregation approaches have been proposed and the
reader is referred to \cite{BS09} for a survey of popular
aggregation methods used in bioinformatics.

Both theoretical and experimental results have suggested that the
generation of a set of \emph{diverse} component learners is one of
the keys to the success of ensemble learning \cite{BWH05}. To
construct diverse local learners, two strategies are widely used:
data perturbation and function perturbation.

\subsubsection{Data Perturbation}

\begin{figure}
\centerline{\includegraphics[width=0.75\textwidth]{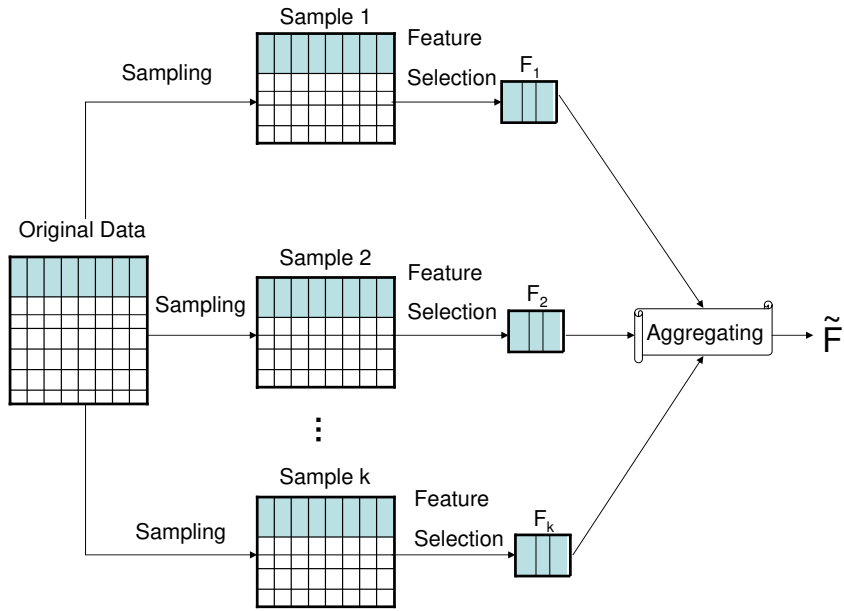}}
\caption{Ensemble feature selection using data perturbation. We
first use different sub-samplings of training data to select
features and then build a consensus output with a rank aggregation
method. } \label{fig:dataPerturbation}
\end{figure}

Data perturbation tries to run component learners with different
sample subsets (e.g., Bagging \cite{Bagging}, Boosting
\cite{Boosting}) or in distinct feature subspaces (e.g., Random
Subspace \cite{Ho98}). In ensemble feature selection with data
perturbation, different samplings of the original data are generated
to construct different feature selectors, as described in
Fig.\ref{fig:dataPerturbation}. Several recent methods
\cite{DGH06,Bach08,MB08,AHP10} fall into this category. These
methods can be further distinguished according to the sampling
method, the component feature selection algorithm and the rank
aggregation method (see Table \ref{tab:dataPerturbation}).

The combination of data sampling and ensemble learning for feature
selection is probably the most intuitive idea to handle selection
instability with respect to sampling variation. The superiority of
such strategy has been verified both experimentally
\cite{DGH06,AHP10} and theoretically \cite{Bach08,MB08}.

Interestingly, all methods listed in Table
\ref{tab:dataPerturbation} are based on the same aggregation scheme,
i.e., linear combination. Note that it is also feasible to combine
data perturbation with other complicated aggregation procedures such
as those ones used in function perturbation (see next subsection).

\begin{table}[h]
\caption{Classification of data perturbation based ensemble feature
selection methods. Here linear combination methods aggregate
rankings using the (weighted) \emph{min}, \emph{max}, or \emph{sum}
operation. The filter method is a general feature selection strategy
that attempts to rank features solely according to their relevance
to target class.} \label{tab:dataPerturbation}
\begin{center}
\begin{tabular}{c c c c}
  \hline
  Reference & Sampling Method & Feature Selector & Aggregation Method \\
  \hline
  Davis et al \cite{DGH06} & Random Subset & Filter Method & Linear Combination  \\
  Bach \cite{Bach08}& Boostrap & Lasso \cite{Lasso} & Linear Combination  \\
  Meinshausen and Buhlmann \cite{MB08}  & Random Subset & Randomized Lasso & Linear Combination  \\
  Abeel et al \cite{AHP10} & Random Subset & SVM-REF \cite{SVMREF} & Linear Combination  \\
  \hline
\end{tabular}
\end{center}
\end{table}

\subsubsection{Function Perturbation}

Here we use function perturbation to refer to those ensemble feature
selection methods in which the component learners are different from
each other. The basic idea is to capitalize on the strengths of
different algorithms to obtain robust feature subsets.

Function perturbation is different from data perturbation in two
perspectives:
\begin{itemize}
  \item It uses different feature selection algorithms rather than
  the same feature selection method.
  \item It typically conducts local feature selection on the original data (without
  sampling).
\end{itemize}

Existing ensemble feature selection methods in this category
\cite{YXS05,DG07,TFR09,NMO09} differ mainly in the aggregation
procedure:
\begin{itemize}
  \item The distance synthesis method is used in \cite{YXS05}.
  \item The Markov chain based rank aggregation method \cite{DKN01} is utilized in \cite{DG07}.
  \item The linear combination method is used in \cite{TFR09}.
  \item The concept of stacking \cite{Stacking} is applied to
  the aggregation of feature selection results in \cite{NMO09}.
\end{itemize}

Compared to data perturbation, function perturbation is less
flexible since the ensemble scale is limited by the number of
available feature selection algorithms in the system. As a result,
no more than four component feature selectors are used in the
ensemble learning process \cite{YXS05,DG07,TFR09,NMO09}.

\subsection{Feature Selection with Prior Feature Relevance}

In most biomarker discovery applications, we typically assume that
all features are equally relevant before the selection procedure. In
practice, some prior knowledge may be available to bias the
selection towards some features assumed to be more relevant
\cite{HD09a,HD09b}. It has been shown that the use of prior
knowledge on relevant features induces a large gain in stability
with improved classification performance \cite{HD09a}.

\begin{figure}
\centerline{\includegraphics[width=0.75\textwidth]{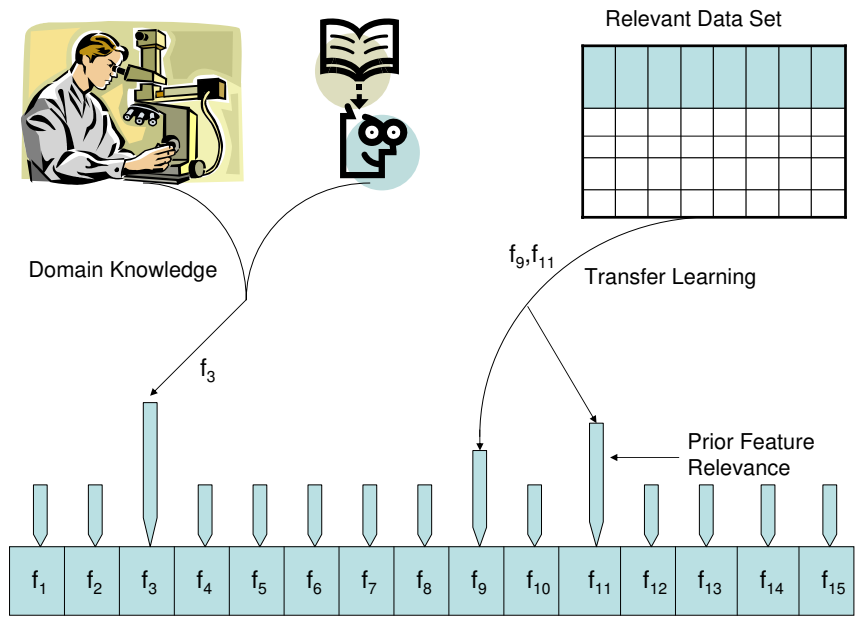}}
\caption{Stable feature selection using prior knowledge on features.
The prior knowledge on relevant features are either obtained from
domain experts or extracted from relevant data sets via transfer
learning.} \label{fig:featurePriori}
\end{figure}

Fig.\ref{fig:featurePriori} shows that there are several methods for
obtaining such kind of prior knowledge. One feasible method is to
seek advices from domain experts or relevant publications. For
instance, in gene expression data classification, one biologist may
know or guess that some genes are likely to be more relevant
\cite{HD09a}.

Another more interesting method is to obtain such prior knowledge
from relevant data sets using transfer learning \cite{HD09b}.
Transfer learning focuses on extracting knowledge from source task
and applying it to a different but related task \cite{PY10}. In
\cite{HD09b}, those features that have been identified as markers
from other data sets are considered to be more relevant in the new
feature selection task.

Though the prior knowledge is helpful in improving the stability of
feature selection, using such information deserves certain
limitations since biomarker discovery aims at finding new features
rather than known ones.

\subsection{Group Feature Selection}

One motivation for group feature selection is that groups of
correlated features commonly exist in high-dimensional data, and
such groups are resistant to the variations of training samples
\cite{LYD09}. If each feature group is considered as a coherent
single entity, potentially we may improve the selection stability.

\begin{figure}
\centerline{\includegraphics[width=0.75\textwidth]{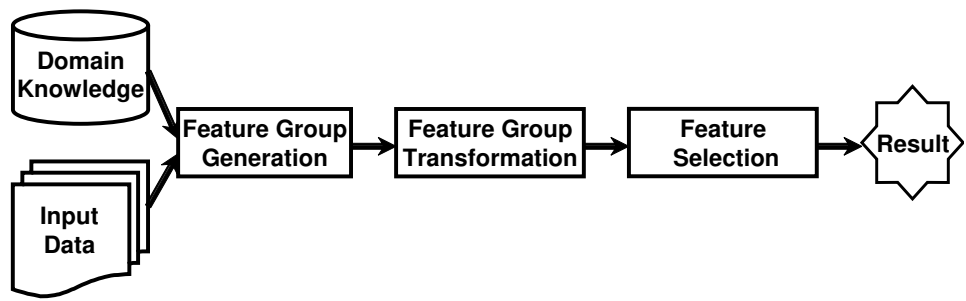}}
\caption{A generic group feature selection framework. In the first
step, we identify feature groups using either knowledge-driven
methods or data-driven approaches. In the second step, we transform
each feature group into a single entity. Finally, we conduct feature
selection in the transformed space.} \label{fig:featureGroup}
\end{figure}

Existing group feature selection algorithms follow the procedure
described in Fig. \ref{fig:featureGroup}. There are two key steps:
group formation and group transformation.

Group formation is the process of identifying groups of associated
features. There are typically two classes of methods for this
purpose: knowledge-driven methods and data-driven methods. The
knowledge-driven group formation method utilizes domain knowledge to
facilitate the generation of feature groups. For example, genes
normally function in co-regulated groups, making it feasible to
search genes in the same pathway for group identification. In
contrast, the data-driven group formation method finds feature
clusters using only information contained in the input data.

Group transformation generates a single coherent representation for
each feature group. The transformation method can range from simple
approaches like feature value mean \cite{GZL05} to complicated
methods such as principal component analysis \cite{RZD07}.

In the following subsections, we will discuss existing group feature
selection methods according to their group formation strategies.

\subsubsection{Knowledge-Driven Group Formation}

\begin{figure}
\centerline{\includegraphics[width=0.75\textwidth]{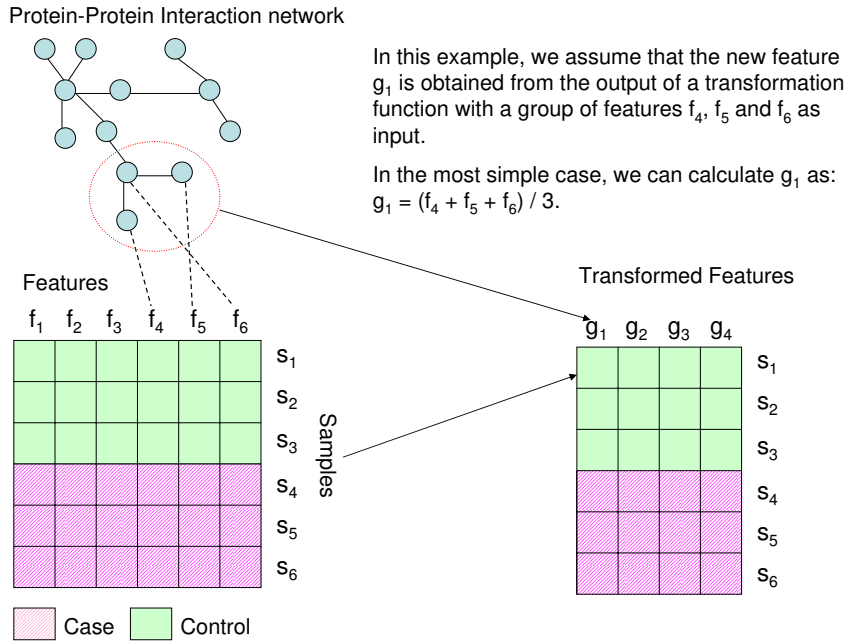}}
\caption{An illustration of knowledge-driven group formation and
feature transformation. The pathway information is used to guide the
search of correlated features (genes or proteins). Each identified
feature group is transformed into a new feature for further
analysis.} \label{fig:pathway}
\end{figure}

Recent advances in the construction of large protein networks make
it feasible to find genes or proteins that have coherent expression
patterns in the same pathway \footnote{For simplicity, this paper
uses the terms ``gene ontology'', ``pathway'', and ``gene set''
interchangeably, although they may not be strictly equivalent.}.
Using available protein-protein interaction (PPI) networks, a number
of approaches have been proposed to incorporate the pathway
information into the biomarker discovery procedure. As shown in
Fig.\ref{fig:pathway}, the basic idea is to find a group of
associated genes or proteins from the same pathway, and then
transform this group into a new entity for subsequent feature
selection and classification. It has been shown that such
knowledge-based method is capable of achieving more reproducible
feature selection and higher accuracy \cite{CLL07}.

We can further distinguish available methods in this category
according to their target data: gene expression data and proteomics
data.

Before summarizing existing approaches on biomarker discovery using
gene pathways, we would like to discuss a closely related problem:
gene set significance testing. Testing the statistical significance
of gene pathways or clusters has been extensively investigated.
Well-known examples include the gene set enrichment analysis
\cite{GSEA} and the maxmean approach \cite{Maxmean}. The reader is
referred to \cite{NK08,DPM09} for comprehensive reviews of existing
approaches on gene set analysis. Here we highlight the fact that
gene set significance testing is different from pathway-guided
biomarker discovery since they have different objectives. The
objective of gene set significance testing is to find whether a
given gene set satisfies the hypothesis, while biomarker discovery
aims at searching for a small subset of genes that can distinguish
cases from controls as accurately as possible. Their intrinsic
connection is that we can utilize pathway significance assessment
method as a filter method (a special type of feature selection
technique that considers each entity individually) for ranking
pathway markers.

Some methods have been proposed to identify markers not as
individual genes but gene sets
\cite{GZL05,RZD07,CLL07,TP07,LCK08,HP09,YKM09,CW09,SYD10}. In Table
\ref{tab:genePathway}, we provide a brief summary of existing
biomarker identification methods that use gene pathway information
as prior knowledge. These methods exploit different strategies for
group generation and transformation. In group generation, we can use
all genes in the pathway for a clear biological interpretation.
Alternatively, we can search for a subset of genes so as to obtain
one more discriminating group. To effectively represent each group,
various summary statistics have been applied, ranging from mean to
principal component analysis.

\begin{table}[h]
\caption{Summary of gene pathway biomarker discovery methods. In the
generation of gene groups, we either accept all genes in a given
pathway or use heuristic search methods (such as greedy algorithm)
to find a subset of discriminating genes. Here ``No transformation''
means that we use all the genes in the group to represent this gene
set. GXNA (Gene eXpression Network Analysis) is a software package
developed in \cite{NTL07} for identifying a subset of differentially
expressed genes from a given pathway.} \label{tab:genePathway}
\begin{center}
\begin{tabular}{c c c}
  \hline
  Reference & Group Generation &Group Transformation \\
  \hline
   Guo et al \cite{GZL05} & Use all genes &Mean and median  \\
   Rapaport et al \cite{RZD07} &Use all genes &Principal component analysis \\
   Chuang et al \cite{CLL07} &Greedy search &Sum of z-scores\\
   Tai and Pan \cite{TP07}  &Use all genes &No transformation \\
   Lee et al \cite{LCK08} &Greedy search &Sum of z-scores \\
   Hwang and Park \cite{HP09}  &Greedy search &Mean  \\
   Yousef et al \cite{YKM09} & GXNA \cite{NTL07} &No transformation  \\
   Chen and Wang \cite{CW09}  &Use all genes &Principal component analysis  \\
   Su et al \cite{SYD10} &Use all genes &Sum of log-likelihood ratio \\
  \hline
\end{tabular}
\end{center}
\end{table}

Recently, such knowledge-driven approach has also been applied to
proteomics data for biomarker discovery at the level of protein
group \cite{JZW08,LNL09}. Compared to gene expression data, more
research efforts toward this direction are desired in future
research.

The pathway-guided group formation method has the advantage that new
transformed features are biologically interpretable since the
underlying disease process may be dependent on perturbations of
different pathways. Thus, prediction models based on pathways may
approximate the true disease process more closely than gene-based
models \cite{CW09}. Its main disadvantage is that we may group
unrelated genes or proteins since the reliability of the predicted
interactions in PPI network is still questionable \cite{MKS02}.

\subsubsection{Data-Driven Group Formation}

Instead of relying on prior knowledge of biology, the data-driven
group formation method identifies feature clusters using either
cluster analysis
\cite{HTB01,JY03,ACW05,MSH07,MH07,YJS07,PHT07,SSJ08,TZH08} or
density estimation \cite{YDL08,LYD09}. As summarized in Table
\ref{tab:clustering}, clustering-based methods utilize popular
partition algorithms such as hierarchical clustering or $k$-means to
generate feature groups. It should be noted that most existing
clustering-based methods do not explicitly consider the stability of
feature group. Alternatively, kernel density estimation is utilized
in \cite{YDL08,LYD09} based on the observation that dense core
regions are stable respect to samplings of dimensions.

\begin{table}[h]
\caption{Summary of clustering-based group feature selection methods
according to clustering algorithms and group transformation methods.
} \label{tab:clustering}
\begin{center}
\begin{tabular}{c c c}
  \hline
  Reference &Clustering Methods &Group Transformation \\
  \hline
   Hastie et al \cite{HTB01} & Hierarchical clustering& Mean \\
   Jornsten and Yu \cite{JY03}  & Integrated clustering and group selection &Mean  \\
   Au et al \cite{ACW05} & $K$-modes &One most discriminating feature \\
   Ma et al \cite{MSH07} & $K$-means &A subset of most discriminating features \\
   Ma and Haung \cite{MH07}  & Hierarchical clustering/$K$-means &A subset of most discriminating features \\
   Yousel et al \cite{YJS07} & $K$-means &No transformation  \\
   Park et al \cite{PHT07} & Hierarchical clustering &Mean \\
   Shin et al \cite{SSJ08} & Hierarchical clustering &One most discriminating feature  \\
   Tang et al \cite{TZH08} & Fuzzy $k$-means &No transformation \\
  \hline
\end{tabular}
\end{center}
\end{table}

There is another class of related methods assigning comparable
coefficients to correlated, important variables. The ``elastic net''
\cite{ZH05} is a typical example in this category. We omit these
methods in this survey since they didn't explicitly identify feature
groups.

The data-driven group feature selection method fully exploits the
characteristics of target data so that it is widely applicable. One
main drawback is that it is not easy to interpret and validate the
selected feature group biologically. One possible remedy is to use a
hybrid strategy that combines the data-driven method with the
knowledge-driven method, as recently discussed in \cite{MS08,MHS09}.

\subsection{Feature Selection with Sample Injection}

In biomarker discovery applications, the number of features is
typically larger than the sample size. This is one of the main
sources of instability in feature selection. To increase the
reproducibility of feature selection, one natural idea is to
generate more samples. However, the generation of real sample data
from patients and healthy people is usually expensive and
time-consuming. With this practical limitation in mind, people begin
to seek other alternative methods for the same purpose.

From the viewpoint of data analysis, there are two data augmentation
strategies:
\begin{itemize}
  \item Utilizing test data to increase the sample size in feature
  selection process, which can be modeled as a transductive learning problem \cite{vapnik98}.
  \item Generating some artificial training samples according to the
  distribution of available training data.
\end{itemize}

In the following sections, we will introduce each method in detail.

\subsubsection{Method Using Transductive Learning}

\begin{figure}
\centerline{\includegraphics[width=0.75\textwidth]{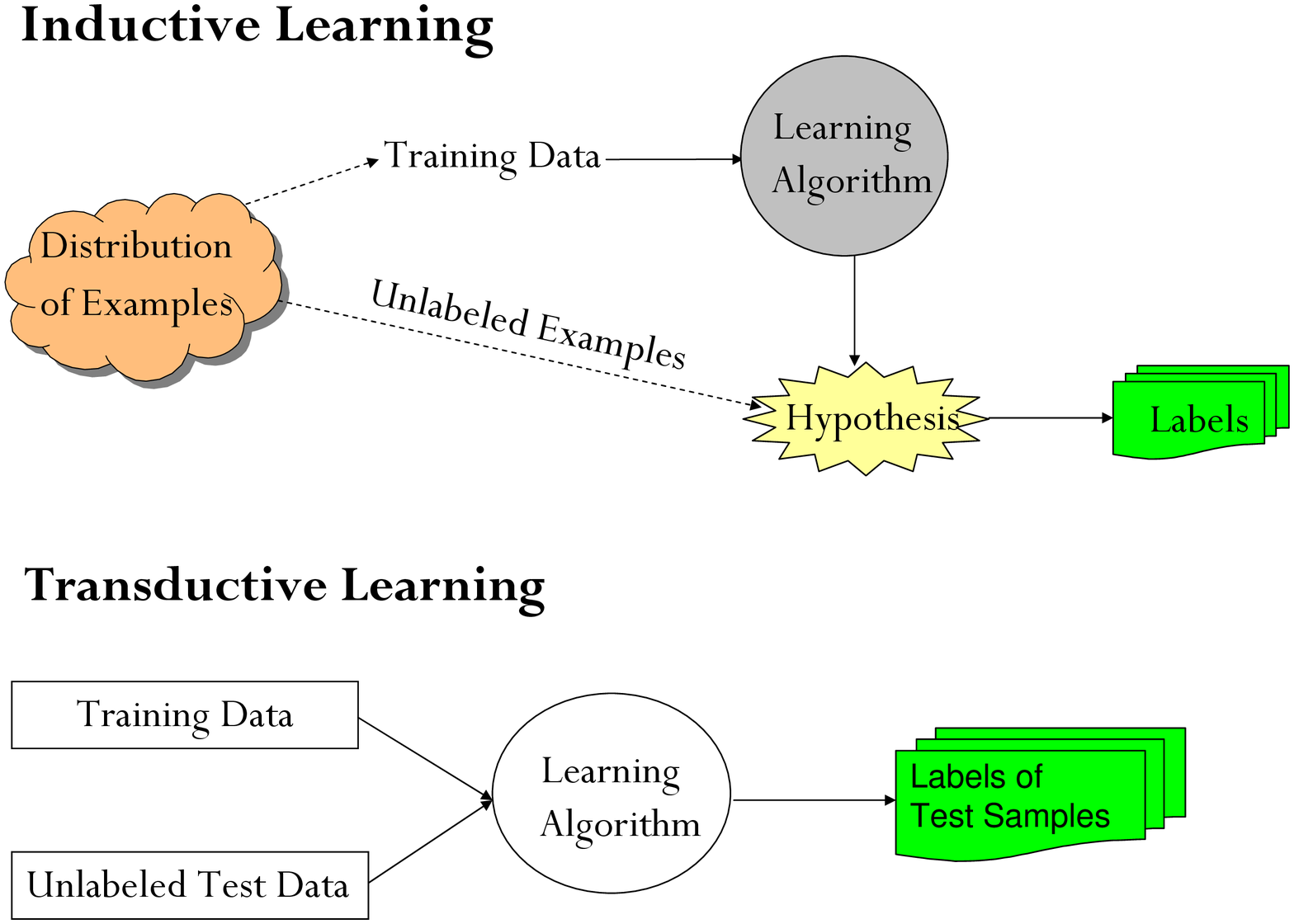}}
\caption{An illustration of inductive learning and transductive
learning. Intuitively, we can consider inductive learning as ``an
education system for all-round development'' while transductive
learning is an ``examination-oriented education system''.}
\label{fig:transductive}
\end{figure}

Different from inductive learning algorithms, the transductive
learning algorithm is not required to produce a general hypothesis
that can predict the label of any unobserved data \cite{vapnik98}.
As illustrated in Fig. \ref{fig:transductive}, it is only required
to predict the labels of a given test set of samples. In other
words, we can use both training data and testing data in the
learning procedure.

Transductive learning has been used to increase sample size in some
recent papers \cite{ZM08,HST08}. The main idea is to take advantage
of the information embedded in the test data so that the role of
test samples is changed from \emph{passive} to \emph{active}. That
is, the unlabeled test samples are incorporated into the feature
selection and classification process.

\subsubsection{Method Using Artificial Training Samples}

\begin{figure}
\centerline{\includegraphics[width=0.75\textwidth]{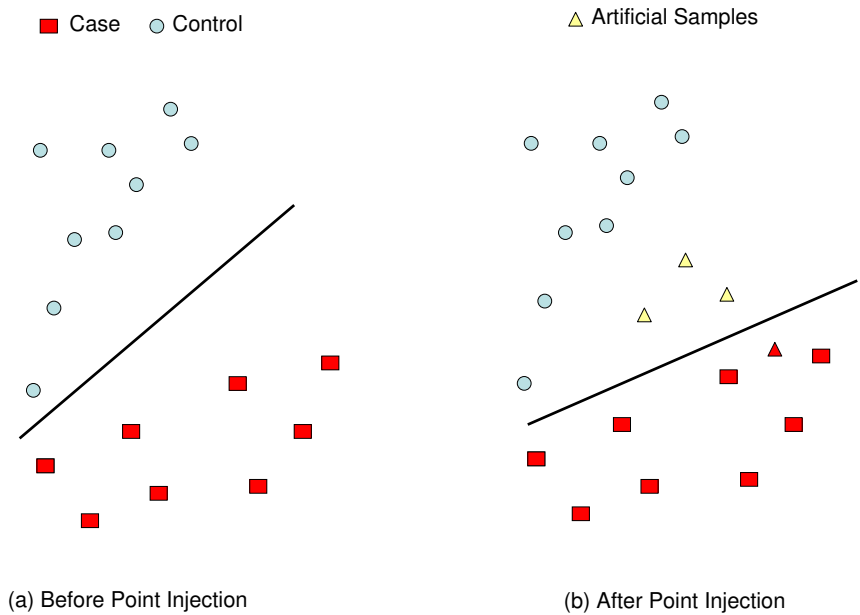}}
\caption{The effect of artificial training samples on model
selection. The separating hyperplane obtained from the data set with
injected samples is different from that of the original samples.}
\label{fig:artificialSample}
\end{figure}

The idea is to generate a number of artificial training samples
according to the distribution of given samples. Then, feature
subsets can be assessed using both the generated data and the
original data. In Fig.\ref{fig:artificialSample}, we provide an
example to illustrate the effect of injected artificial samples on
model selection.

There are many methods for generating artificial training samples.
For instance, we can first pick one training sample $x_{i}$ randomly
and then generate a point $z$ from standard normal distribution.
Finally, we generate the new artificial point as: $y=x_{i}+hz$,
where $h$ is a constant.

There are two ways in which the artificial points participate in
feature selection. One method is to treat the injected points as the
original samples in the training process \cite{KDB02}. Another
method is to use the injected points only in the evaluation stage
\cite{HC07}.

\section{Stability Measure}

In stable feature selection, one important issue is how to measure
the ``stability'' of feature selection algorithms, i.e., how to
qualify the selection sensitivity to variations in the training set.
The stability measure can be used in different contexts. On the one
hand, it is indispensable for evaluating different algorithms in
performance comparison. On the other hand, it can be used for
internal validation in feature selection algorithms that take into
account stability.

Noticing that there is already a nice review paper \cite{BS09} on
the stability of ranked gene lists, here we would like to provide a
more comprehensive list that includes evaluation methods from
different domains.

Measuring stability requires a similarity measure for feature
selection results. This depends on the representations used by
different algorithms. Formally, let training examples be described
by a vector of features $F = (f_{1}, f_{2},..., f_{m})$, then there
are three types of representation methods \cite{KPH07}:

\begin{itemize}
  \item A subset of features: $S=\{s_{1},s_{2},..,s_{k}\}$, $s_{i}\in
  \{f_{1}, f_{2},..., f_{m}\}$.
  \item A ranking vector: $R=(r_{1},r_{2},..,r_{m})$, $1\leq r_{i}\leq m$.
  \item A weighting-score vector: $W=(w_{1},w_{2},..,w_{m})$, $w_{i}\in
  R^{+}$.
\end{itemize}

In general, we are interested in stability measures that take more
than two subsets (or rankings) into account. In this review, we use
measures defined on two subsets (or rankings) for the sake of
notation simplification. As pointed out in \cite{BS09}, there are
essentially two approaches for generalizing the definition. One
approach is to summarize pairwise stability measures through
averaging. Another approach is to consider all subsets (or rankings)
simultaneously in the specification of stability measure.

In the following, we will summarize available stability measures
according to the representation of feature selection results.

\subsection{Feature Subset}

There is a wide variety of similarity measures available for the
comparison of sets. Table \ref{tab:measureSubset} summarizes
available stability measures. One may find that most of these
measures are defined using the physical properties of two sets,
e.g., the ratio of the intersection to the union. One exception is
the ``percentage of overlapping features related'' \cite{ZZZ09},
which incorporates additional feature correlation information into
the measure definition. This is definitely plausible for biomarker
discovery applications since there are always some highly correlated
features in the ``omics'' data.

\begin{table}[h]
\caption{A list of stability measures when the feature selection
algorithm produces a subset of features as output. Here $S$ and
$S^{'}$ are two subsets of features. } \label{tab:measureSubset}
\begin{center}
\begin{tabular}{c c c c}
  \hline
  Index &Description &Formula  \\
  \hline
  MS1 \cite{DCA02} &Relative Hamming distance& $1-\frac{|S\setminus S^{'}|+|S^{'}\setminus S|}{m}$  \\
  MS2 \cite{KPH07,Stolovitzky03,NBT09}&Tanimoto distance/Jaccard's index& $1-\frac{|S|+|S^{'}|-2|S\cap S^{'}|}{|S|+|S^{'}|-|S\cap S^{'}|}$  \\
  MS3 \cite{YDL08,ZRS08,LYD09}&Dice-Sorensen's index & $\frac{2|S\cap S^{'}|}{|S|+|S^{'}|}$  \\
  MS4 \cite{ZRS08} &Ochiai's index & $\frac{2|S\cap S^{'}|}{\sqrt{|S||S^{'}|}}$ \\
  MS5 \cite{Shi05} &Percentage of overlapping features & $\frac{|S\cap S^{'}|}{|S|}$, $\frac{|S\cap S^{'}|}{|S^{'}|}$ \\
  MS6 \cite{ZZZ09}&Percentage of overlapping features related& $\frac{|S\cap S^{'}|+c_{12}}{|S|}$, $\frac{|S\cap S^{'}|+c_{21}}{|S^{'}|}$ \\
  MS7 \cite{Kuncheva07}&Kuncheva's stability measure & $\frac{|S\cap S^{'}|m-c^{2}}{c(m-c)}$  \\
  MS8 \cite{SN08} &Consistency & $\frac{1}{|S\cup S^{'}|}\underset{f\in S\cup S^{'}}\sum \frac{freq(f)-1}{m-1}$ \\
  MS9 \cite{SN08}&Weighted consistency & $\underset{f\in S\cup S^{'}}\sum(\frac{freq(f)}{|S|+|S^{'}|}\cdot\frac{freq(f)-1}{m-1})$  \\
  MS10 \cite{DGH06} &Length adjusted stability & $max\{0,\underset{f\in S\cup S^{'}}\sum(\frac{freq(f)}{2|S\cup S^{'}|}-\alpha \frac{|S|+|S^{'}|}{2m})\}$ \\
  \hline
\end{tabular}
\end{center}
\end{table}

\textbf{MS1}: The relative Hamming distance between the masks
corresponding to two subsets is used to measure the stability
\cite{DCA02}.

\textbf{MS2}: The Tanimoto distance metric measures the amount of
overlap between two sets of arbitrary cardinality. It takes values
in [0,1], with 0 meaning no overlap between the two sets and 1
meaning two sets are identical. In fact, this measure is equivalent
to the Jaccard's index: $\frac{|S\cap S^{'}|}{|S \cup S^{'}|}$.

\textbf{MS3}: The Dice-Sorensen's index is the harmonic mean of
$\frac{|S\cap S^{'}|}{|S|}$ and $\frac{|S\cap S^{'}|}{|S^{'}|}$.

\textbf{MS4}: The Ochiai's index is the geometric mean of
$\frac{|S\cap S^{'}|}{|S|}$ and $\frac{|S\cap S^{'}|}{|S^{'}|}$. It
has been shown that the performance of the Ochiai's index is similar
with that of Jaccard's index and Dice-Sorensen's index \cite{ZRS08}.

\textbf{MS5}: This measure is originally named as: ``Percentage of
Overlapping Genes (POG)'' in the context of gene expression data
analysis.

\textbf{MS6}: It is an extension of POG, which incorporates highly
correlated features between two sets into the stability evaluation.
In the formula, $c_{12}$ (or $c_{21}$) denotes the number of
features in $S$ (or $S^{'}$) that are not shared but are
significantly positively correlated with at least one feature in
$S^{'}$ (or $S$). The normalized form of this measure is also
presented in \cite{ZZZ09}.

\textbf{MS7}: This stability measure assumes that $S$ and $S^{'}$
have the same size (cardinality), i.e., $|S|=|S^{'}|=c$.

\textbf{MS8 and MS9}: In both definitions, $freq(f)$ denotes the
number of occurrences (frequency) of feature $f$ in $S\cup S^{'}$.
It has been proved that both measures take values in [0,1]. The
(weighted) consistency value is 1 if two sets are identical and 0 if
they are disjoint.

\textbf{MS10}: In the formula, $\alpha$ is one user-specified
parameter and is set to 10 \cite{DGH06}. Note that $(|S|+|S^{'}|)/2$
corresponds to the median required in \cite{DGH06} since there are
only two sets in our formulation.

\subsection{Ranking List}

The problem of comparing ranking lists is widely studied in
different contexts such as voting theory and web document retrieval.
Table \ref{tab:measureRank} shows some distance measures for two
ranking lists. One typical example is MR2, in which the Spearman's
correlation is adapted to place more weights on those top ranked
features since these features are more important than irrelevant
features in the stability evaluation.

\begin{table}[h]
\caption{A list of stability measures when the feature selection
algorithm generates a ranking list as output. Here $R$ and $R^{'}$
are two different ranking lists. } \label{tab:measureRank}
\begin{center}
\begin{tabular}{c c c c}
  \hline
  Index &Description &Formula \\
  \hline
  MR1 \cite{KPH07} &Spearman's rank correlation coefficient & $1-6\underset{i=1}{\overset{m}\sum}\frac{(r_{i}-r^{'}_{i})^{2}}{m(m^{2}-1)}$   \\
  MR2 \cite{JMB08} &Canberra distance & $\underset{i=1}{\overset{m}\sum}\frac{|min\{r_{i},k+1\}-min\{r_{i}^{'},k+1\}|}{min\{r_{i},k+1\}+min\{r_{i}^{'},k+1\}}$ \\
  MR3 \cite{YBS06} &Overlap score & $\underset{i=1}{\overset{m}\sum}e^{-\alpha i}|\{r_{j}|j<i\}\cap\{r_{j}^{'}|j<i\}|$  \\
  \hline
\end{tabular}
\end{center}
\end{table}

\textbf{MR1}: The Spearman's rank correlation coefficient takes
values in [-1,1], with 1 meaning that the two ranking lists are
identical and a value of -1 meaning that they have exactly inverse
orders.

\textbf{MR2}: The Canberra distance is a weighted version of
Spearman¡¯s footrule distance \cite{JMB08}, i.e.,
$\underset{i=1}{\overset{m}\sum}\frac{|r_{i}-r^{'}_{i}|}{r_{i}+r^{'}_{i}}$.
Since the most important features are usually located at the top of
the ranking list \cite{JMB08}, the distance calculation in the table
only considers top $k$ ranked features.

\textbf{MR3}: The overlap score is originally proposed in
\cite{YBS06} and here we follow \cite{BS09} to reformulate it with
the assumption that only top ranked features are important. In the
formula, $\alpha$ is a user-specified parameter to control the
decreasing rate.

\subsection{Weighting-Score Vector}

The computational issue of combinatorial search for feature subset
can to some extent be alleviated by using a feature weighting
strategy \cite{IterativeRELIEF}. Allowing feature weights to take
real-valued numbers instead of binary ones enables us to use
well-established optimization techniques in algorithmic development.
For instance, the RELIEF algorithm \cite{RELIEF} is one
representative of such kind of methods, which generates a weighting
score vector as output.

The weighting score vector is seldom used in defining stability
measure. Table \ref{tab:measureWeight} lists one stability measure
MW1 \cite{KPH07}. The Pearson's correlation coefficient ranges from
-1 to 1. A value of 1 indicates a perfect positive correlation, a
value of 0 means that there is no correlation, while a value of -1
means that they are anti-correlated. In the formula, $u_{W}$ and
$u_{W^{'}}$ are the means of weight scores of $W$ and $W^{'}$,
respectively.

\begin{table}[h]
\caption{The Pearson's correlation coefficient measure. Here $W$ and
$W^{'}$ are two different weighting score vectors. }
\label{tab:measureWeight}
\begin{center}
\begin{tabular}{c c c c}
  \hline
  Index &Description &Formula \\
  \hline
  MW1 \cite{KPH07} &Pearson's correlation coefficient& $\frac{\sum_{i}(w_{i}-u_{W})(w^{'}_{i}-u_{W^{'}})}{\sqrt{\sum_{i}(w_{i}-u_{W})^{2}\sum_{i}(w^{'}_{i}-u_{W^{'}})^{2}}}$ \\
  \hline
\end{tabular}
\end{center}
\end{table}

\section{Discussions}

We summarize three sources of instability for feature selection in
section 2. Among these sources, probably the small number of samples
in high dimensional feature space is the most difficult one in
biomarker discovery. Besides feature selection, other data analysis
tasks also face the same challenges. Research progresses in related
fields will facilitate the development of effective stable feature
selection methods as well.

Group feature selection is the most extensively studied method among
existing stable feature selection approaches. This is because there
are many correlated features in high dimensional space. However,
such feature grouping strategy can only partially alleviate
selection instability since we still need to face the
reproducibility issue in the transformed space. In this regard,
ensemble feature selection is probably more promising to provide a
general-purpose solution. One immediate hybrid strategy is to
combine group feature selection with ensemble feature selection,
i.e., first perform feature grouping and then use ensemble feature
selection in the new feature space.

The group feature selection strategy is only helpful when multiple
sets of true markers are generated due to the existence of redundant
features. However, it is also possible that multiple sets of true
markers share no correlated features. The feature selection problem
in this case is much harder than finding a minimal optimal feature
set for classification \cite{NPB07}. To our knowledge, there is
still no available method and measure that aim at handling stability
issues in this context. The general problem is open and needs more
research efforts.

With respect to stability index, most available measures are defined
over feature subsets since the feature subset can be obtained from
rankings or scores (but not vice-versa). The major problem is that
there is still no consensus on the best stability measure.
Therefore, a comprehensive comparison study on existing stability
measures should be conducted in future research.

In fact, the biomarker discovery process involves many procedures.
Here we only discuss feature selection techniques for stable
biomarker identification. The development of biomarker classifier is
also very important. The readers are referred to a recent review
\cite{BTC09} for research progress towards this direction.

Finally, we would like to raise the following questions in the
pursuit of stable biomarker discovery methods for future research:
\begin{itemize}
  \item How to directly measure the stability of feature(s) without sampling training data?
  \item Can we propose new methods that are capable of explicitly controlling the stability of reported feature subset?
  \item Are there other special requirements for biomarker discovery rather than stability?
\end{itemize}

\section{Conclusions}

To discover reproducible markers from ``omics'' data, the stability
issue of feature selection has received much attention recently.
This review summarizes existing stable feature selection methods and
stability measures. Stable feature selection is a very important
research problem, from both theoretical perspective and practical
aspect. More research efforts should be devoted to this challenging
topic.

\section*{Acknowledgments}
This work was supported with the general research fund 621707 from
the Hong Kong Research Grant Council, a research proposal
competition award RPC07/08.EG25 and a postdoctoral fellowship from
the Hong Kong University of Science and Technology.

\bibliographystyle{IEEEtran}
\bibliography{SFSBD}

\end{document}